\documentclass[fleqn,twoside]{article}
\usepackage{espcrc2}
\usepackage{epsfig}

\usepackage{graphicx}
\usepackage[figuresright]{rotating}

\newcommand{\AmS}{{\protect\the\textfont2
  A\kern-.1667em\lower.5ex\hbox{M}\kern-.125emS}}

\title{Manifestation of Confinement in the Gluon propagator}

\author{K. Langfeld\address[ITU]{Institut f\"ur Theoretische Physik, 
        Universit\"at T\"ubingen, Auf der Morgenstelle 14, \\ 
        72076 T\"ubingen, Germany }\thanks{supported by {\tt Strukturfond 
              2000 of the Uni\-versity of T\"ubingen.} }
        H. Reinhardt\addressmark[ITU]\thanks{talk presented by HR; supported 
           by DFG RE856/4-1} and 
        J. Gattnar\addressmark[ITU] }
       
\newcommand{\tr}{\hbox{tr}}

\newcommand{\no}{\noindent}
\newcommand{\be}{\begin{eqnarray}}
\newcommand{\ee}{\end{eqnarray}}

\begin{document}

\begin{abstract}
The gluon propagator in Landau gauge is calculated on the lattice for SU(2)
gauge theory. In particular the manifestation of confinement in the gluon
propagator is studied. Removing the confining center vortices from the
Yang-Mills ensemble leads to a drastic reduction of the gluonic form
factor in the intermediate momentum regime. 
\vspace{1pc}
\end{abstract}

\maketitle

\no
\section{INTRODUCTION} 

Understanding the physics of hadrons in particular the 
mechanism of confinement
and spontaneous breaking of chiral symmetry requires a non-perturbative
treatment of QCD, which is the theory of strong interaction. To date the only
rigorous non-perturbative approach to QCD are the Monte Carlo lattice
calculations. Despite recent progress by improved lattice algorithm and the
increase in computer time, lattice calculations including dynamical quarks are
exeedingly cumbersome and finite baryon densities are hardly accessible in 
realistic SU(3) lattice simulations. 
\medskip

Dynamical fermions and finite baryon densities can be relatively easily treated in
the Schwinger-Dyson-equation approach to continiuum QCD \cite{Alk}, which is also non-perturbative, 
but less rigorous in the sense that truncations of the tower of 
coupled equations are necessary in practical calculations, which however are not well
under control. Despite this shortcoming the Schwinger-Dyson-approach has
been successfully applied to the description of hadron 
phenomenology~\cite{Rob}. The central
quantity of interest in this approach is the gluon propagator. It is
therefore of primary interest to obtain rigorous results on the gluon
propagator, which to date can be obtained only by lattice Monte Carlo
calculations. Of particular interest is how confinement manifests itself in the
gluon propagator. This is the main subject of the present study. 
\medskip

Since the gluon propagator is gauge dependent, its properties can be specified
only after fixing the gauge. In the Schwinger-Dyson-approach it is common 
to use the Landau gauge 
\be 
\partial_\mu A_\mu(x)=0 \; , 
\label{G1} 
\ee 
and the gluon propagator has been studied in this gauge on the lattice in 
\cite{Cuc}. Here we studied the signature of quark confinement which 
might be encoded in the gluon propagator in this gauge. 
\medskip

Recent lattice calculations give strong support for the vortex picture of
confinement according to which the ``confiner'' are center vortices \cite{DeD}. We will
therefore investigate the effect of center vortices on the gluon propagator.

\no
\section{LATTICE DEFINITION OF THE GLUON PROPAGATOR } 

Lattice gauge theory is defined in terms of link variables $U_\mu(x)$ from 
which the gauge field $A_\mu(x)$ of the continiuum theory is usually 
defined by $U_\mu(x)=\exp\{- a \, A^b_\mu(x)T^b\}$ 
where $a$ is the lattice spacing and $T^b$ denotes the generators of the gauge
group in the fundamental representation. Note that the so defined gauge field
$A^a_\mu(x)$ becomes singular when $U_\mu(x) \rightarrow -1.$ 
Furthermore under gauge transformations it does not properly transform as a
connection of the continuum theory. In particular, it is not invariant under
center gauge transformations. 
Therefore we prefer to perform a coset decomposition of the links
\be
\label{G3}U_\mu(x)=Z_ \mu(x)\bar{U}_\mu(x), \, 
Z_\mu(x)= \mathrm{sign} \, \tr U_\mu(x),
\ee
where $Z_\mu (x) \in \{1, -1\}$ denotes the center element closest to $U_\mu(x)$, and
define the gauge field $A_\mu (x)$ from the coset part $\bar{U}_\mu(x)$ or equivalently from the
links in the adjoint representation 
\be
\label{G4}\hat{U} _\mu (x)^{ab} & = &-2 \tr 
\bigl\{ U_\mu (x) T^a  U_ \mu ^\dagger (x)T^b \bigr\} 
\nonumber \\
& = & \exp \bigl\{ -a A_\mu ^c (x) \hat{T}^c \bigr\} \; , 
\ee
where $(\hat{T}_c)^{ab} = f^{acb}$ are the structure constants. 
\medskip

With the continuum gauge field $A_\mu^a(x)$ at hand the gluon propagator is
given by
\be
\label{G5}D^{ab}_{\mu \nu}(x-y)=\big<A^a_\mu(x)A^b_\nu(y)\big>
\ee 
where $<...>$ denotes the Monte Carlo average over properly thermalized lattice
gauge configurations.  
The Fourier transform of (\ref{G5}) is expressed
in terms of the lattice momentum $p_\mu=\frac{2}{a(\beta)} 
\sin \frac {n_\mu\pi}{N_\mu}$ which reduces to the Matsubara
frequency $\bar{p}_\mu=\frac {2\pi}{N_\mu \, a(\beta ) } n_\mu$ for 
$n_\mu\ll N_\mu  $. In this
variable the free lattice gluon propagator has the form
$D_0 (p)=\frac{1}{\bar{p}^2}$. In the following we will be interested in the 
non-perturbative information contained in the gluon propagator $D(p)$, which is
measured by the form factor $F(p)=p^2D(p)$. High precision Monte Carlo
measurements of the form factor are obtained by extending the method of
\cite{Cuc} by choosing a purely temporal momentum transfer $\bar{p}=(0, 0, 0,
\bar{p}_4)$ and expressing the gluonic form factor as

\be
\label{G7}F(\bar{p}_4^2) &=& \sum_{a,_\mu}\frac{1}{N^2} \big< \Big[ \sum_x
\Delta_t A^a_\mu (x) \, \cos \, \bar{p}x\Big]^2 
\nonumber\\
&+& \Big[\sum_y \Delta_t A^a_\mu (y) \, 
\sin \, \bar{p} y \Big]^2\big>
\ee 

where $\Delta_t A(x)=A(x + a\vec{e}_4)-A(x)$. 
\medskip

The gluon propagator is gauge dependent and is properly defined only after
fixing the gauge. In fact in the unfixed gauge lattice theory the gluon
propagator vanishes by center symmetry. In order to be able to compare with the
Schwinger-Dyson-approach to continuum Yang-Mills-theory we use the (lattice)
Landau gauge $\sum_{x,\mu} tr U_\mu(x) \rightarrow max $. 
\begin{figure}[t]
\centerline{
\epsfxsize=\linewidth 
\epsfbox{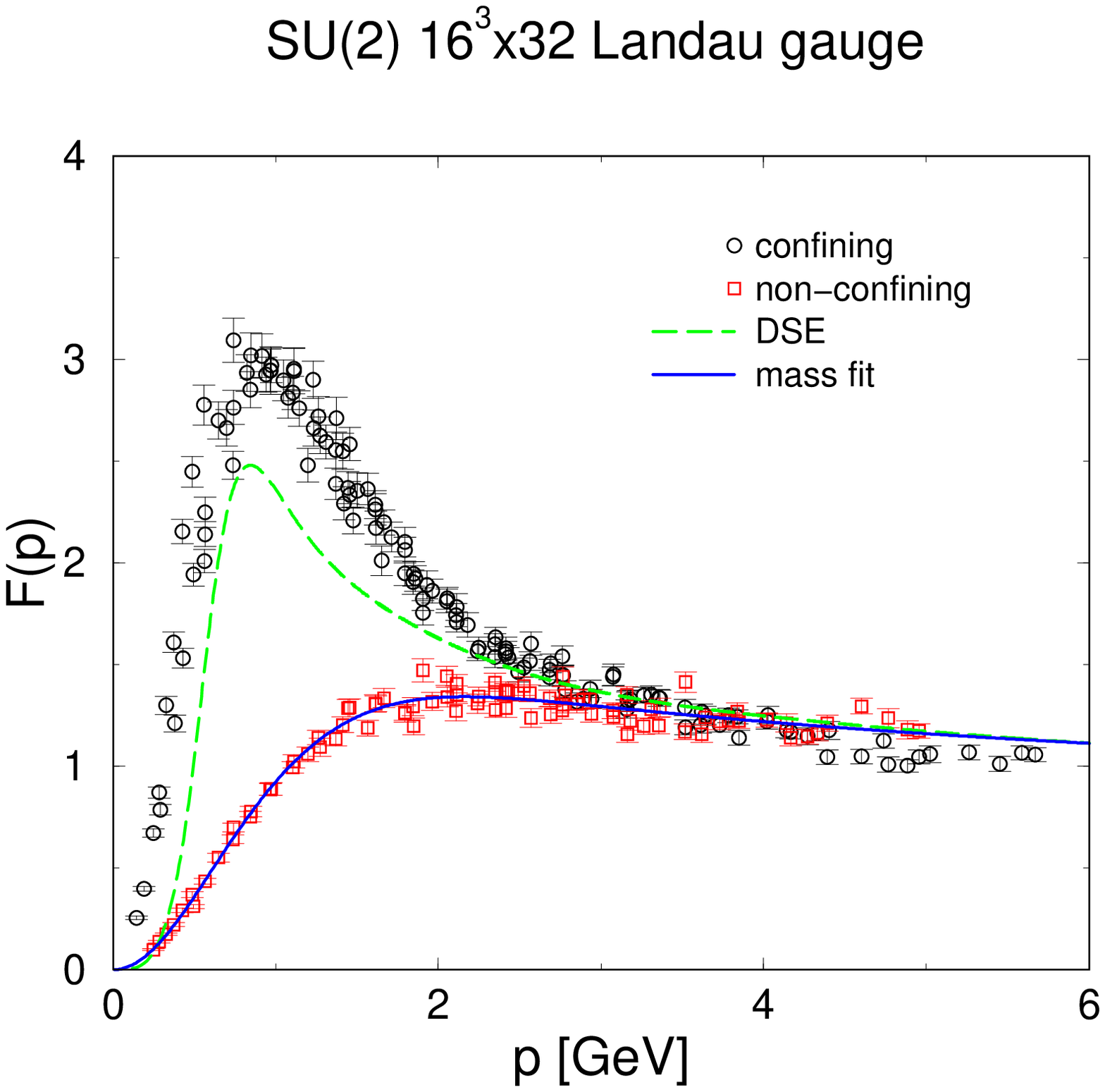}
}
\label{fig:1}
\end{figure}
which reduces in the continuum limit $a \rightarrow 0$ to the familiar Landau
gauge, eq. (\ref{G1}). The lattice Landau gauge exploits the gauge freedom to bring each
link as close as possible to the unit matrix. Consequently in this gauge most
of the center elements in the coset decomposition,  eq. (\ref{G3}), are trivial
$(Z_\mu(x)=1)$ and most of the physics is transferred to the adjoint links. 
Fig.~1 shows the gluon form factor in the Landau gauge
evaluated from eq.~(\ref{G7}). The gluon form factor vanishes at zero 
momentum transfer $q=0$, has a
rather pronounced peak at intermediate momenta and asymptotically 
approaches the perturbative tail. 
Also shown in Figure 1 is the gluon propagator obtained by
solving the coupled gluon-ghost Schwinger-Dyson-equations of continuum
Yang-Mills-theory in the Landau gauge, eq. (\ref{G1}),~\cite{chfi}. The
Schwinger-Dyson-approach qualitatively reproduces the lattice gluon propagator,
but misses a substantial part of the peak at intermediate momenta. This missing
strength of the Schwinger-Dyson gluon propagator could originate from: 
truncation of the Schwinger-Dyson equations or approximations in the 
numerical solutions of the coupled Schwinger-Dyson
equations (e.g.~angle approximation). 

\no
\section{MANIFESTATION OF CONFINEMENT IN THE GLUON PROPAGATOR } 

Let us now 
investigate how confinement manifests itself in the gluon propagator.
Lattice calculations have given strong support for the center vortex picture of
confinement. The confining center vortices can be easily detected (and removed)
by the method of center projection on top of the maximum center gauge fixing.
The maximum center gauge \cite{DeD} is equivalent to the adjoint Landau gauge, which
brings the adjoint links as close as possible to the unit matrix 
$$
\sum_{x,y} \; \tr _A \hat{U}_\mu(x) \rightarrow max \; . 
$$
Since in this gauge the center elements $Z_\mu=\pm1$ are treated on 
equal footing, a
substantial portion of the center elements are non trivial $(Z_\mu=-1)$. 
It is observed that the confinement physics 
is concentrated on the center degrees of freedom, or
equivalently, on the center vortices\footnote{Center vortices are the only non-trivial field configurations of a
Z(2) theory.}. Indeed performing in this gauge a so-called center projection
replacing each (gauged) link $U_\mu (x)$ by its ``nearest'' center element 
$
Z_\mu (x) = \mathrm{sign} \,  \tr U_\mu (x) 
$ 
results in a center vortex ensemble, which reproduces the
full string tension. On the other hand, 
removing the center vortices by replacing  
\be
\label{G11}U_\mu (x) \rightarrow Z_\mu (x) \; U_\mu (x) 
\ee
eliminates the confining properties, resulting in a vanishing
string tension \cite{Foc}. 
In the continuum limit, the maximal center gauge reduces 
to the background gauge $[ \partial _{\mu}+ V_{\mu}, A_{\mu}] = 0$ 
where $V_{\mu} (x)$ is an ``optimally chosen'' center vortex field
\cite{Eng}. 
Hence after
the center vortices have been removed by the replacement eq.(\ref{G11}) 
the maximum
center gauge reduces in the continuum theory to the Landau gauge eq.(\ref{G1}). The
gluon propagator in the Landau gauge obtained from the non-confining vortex-free
ensemble is also shown in Fig. 1. 
Removal of the confining center
vortices has basically removed the peak at intermediate momenta.
In view of this result we can conclude, that parts of the confining
properties escape the Schwinger-Dyson-approach, at least within the
approximations used in present state of the art calculations. 

\medskip
It would be interesting to solve the Schwinger-Dyson equation for the quark
propagator using the gluon propagator obtained above in the lattice theory as
input and calculate the quark condensate. Using the gluon propagator obtained
in the non-confining theory, where the center vortices have been removed we
expect that the quark condensate $<\bar{q}q>$ vanishes as lattice calculations
indicate \cite{Foc}. 

\medskip
Further details of the work outlined in this talk can be found 
in~\cite{Lan}.

\end{document}